\documentstyle[11pt,epsfig,afterpage]{article}
\def\ifscience{n}
\def\figsin{y}\def\ifdontembed{n}

\def\itppt{\ifx\ifscience\figtest\else\it\fi}

\def\tauchoice{7}

\def\offsetthreed{$-\kappa f(\spb,\stb)/1000\hat n\kbt+218.0+6.0\sbp-11.4\stb$}

\newcommand{\beq}{\begin{equation}}
\newcommand{\eeq}{\end{equation}}

\def\figtest{y}
\newcommand{\pitemize}{\begin{itemize}
\setlength{\itemsep}{0pt}\setlength{\parsep}{0pt}}
\def\tfig#1{Fig.~\ref{#1}}
\def\ifigure#1#2#3#4{\ifx\figsin\figtest
\begin{figure}[t]\begin{center}{\epsfysize=#4truein \epsfbox{#3}}
\end{center}\smallskip
\caption{{\footnotesize #2}\label{#1}}\end{figure}
\else
\begin{figure}[h!]
\caption{\footnotesize #2\label{#1}}\end{figure}
\fi}
\def\kbt{k_{\rm B}T}

\newcommand{\inv}{^{\raise.15ex\hbox{${\scriptscriptstyle -}$}\kern-.05em 1}}

\newcommand{\nhat}{\hat n}

\newcommand{\sbp}{\bar\sigma_+}

\newcommand{\smax}{\sigma_{\rm max}}
\newcommand{\smb}{\bar\sigma_-}\newcommand{\spb}{\bar\sigma_+}
\newcommand{\sob}{\bar\sigma_1}\newcommand{\stb}{\bar\sigma_{\rm t}}
\newcommand{\fm}{f_{\rm m}}
\newcommand{\spav}{\sigma_{+,{\rm av}}}\newcommand{\spavb}{\bar\spav}

\begin{document}

\begin{titlepage}
\begin{center}{\huge
Electrostatic Repulsion of Positively Charged Vesicles and
Negatively Charged Objects}\\
\bigskip
{\small Helim Aranda-Espinoza$^{1}$, Yi Chen$^2$, Nily Dan$^1$,\\
T. C. Lubensky$^2$,  Philip
Nelson$^2$, Laurence Ramos$^3$, and D. A. Weitz$^{2,4}$}\\
\end{center}
\medskip
\begin{flushleft}
{\small$^1$Department of Chemical Engineering, University of Delaware, Newark
DE 19716 USA. Address after 1 Sep 1999: Department of Chemical
Engineering, Drexel University, Philadelphia PA 19104 USA \\
$^2$Department of Physics and Astronomy, University of Pennsylvania,
Philadelphia PA 19104 USA\\
$^3$Groupe de dynamique des phases condens\'ees,
Case 26, Universit\'e de Montpellier II, Place E. Bataillon, 34095
Montpellier Cedex 05, France\\
$^4$Address after 1 Sep 1999: Department of Physics, Harvard
University, Cambridge MA 02138 USA\\
}\end{flushleft}
\bigskip
\begin{center}
{\sl Science 1999, in press}\\
{\sl 11 March 99; revised 11 June 99}\\
\end{center}

\end{titlepage}
\ifx\ifscience\figtest
\stepcounter{page}
\fi
{\bf\noindent
A positively charged, mixed bilayer vesicle in the presence of
negatively charged
surfaces (for example, colloidal particles) can spontaneously partition into an
adhesion zone of definite area, and another zone that {\itppt repels}
additional
negative objects. Although the membrane itself has nonnegative
charge in the repulsive zone, negative counterions on the {\itppt interior} of
the vesicle spontaneously aggregate there, and present a net negative
charge to the exterior. Beyond the fundamental result that
oppositely charged objects can repel, our mechanism helps explain recent
experiments on surfactant vesicles%
.}
\ifx\ifscience\figtest\renewcommand{\section}[1]{}
\renewcommand{\subsection}[1]{}\fi

\section{{Introduction}}
Opposite charges attract in vacuum. Two ionizable objects in an
electrolyte such as water form a more
complex system, but nevertheless, in many situations a simple rule of
thumb applies: As two planar surfaces approach from infinity they
initially attract if oppositely charged, with a {\itppt screened} Coulomb
potential \cite{safranbook}.

The analysis of this \ifx\ifscience\figtest report\else paper\fi{} was
motivated by two sets of experimental
observations that defy the familiar rule above
\cite{nard98a,ramo1}. Bilayer vesicles were prepared from a
mixture of cationic (positively charged) and neutral
surfactants. In one case, vesicles were
allowed to adhere to a negatively charged substrate
\cite{nard98a}, while in the other,  negatively charged
colloidal particles were introduced into suspensions of vesicles and
the resulting self-assembled structures monitored \cite{ramo1}. The
puzzling observation was that despite the high charge on the vesicles,
they were not uniformly attractive to the particles or surfaces, but
instead separated macroscopically into adhesive and non-adhesive zones
(\tfig{fphoto}). The vesicle diameter was typically 20$\,\mu$m; the
Debye screening length $1/\kappa$ was much smaller, between 1 and
10~nm \cite{fnY}. Because membranes in living cells also include bilayers made
from mixtures of negatively charged and neutral lipids \cite{genn89a},
phenomena like the ones reported here might occur generally.

\def\capphoto{A self-limited array of latex spheres on a
charged surfactant vesicle at room temperature (see \cite{ramo1}). The
spheres are on the surface of a vesicle viewed in differential interference
microscopy, focusing on the array itself (left) or on the vesicle
equator (right). The membrane is a mixture of cationic and neutral surfactants
(DDAB and Triton X, respectively, with octanol added to stabilize
bilayer structures). The sphere diameter is 1$\,\mu$m; the sphere volume
fraction is 0.004.}

\ifx\ifdontembed\figtest\else
\ifx\figsin\figtest
\begin{figure}[h!]
\begin{center}
\begin{minipage}{3.1truein}
{\epsfysize=3.1truein\epsfbox{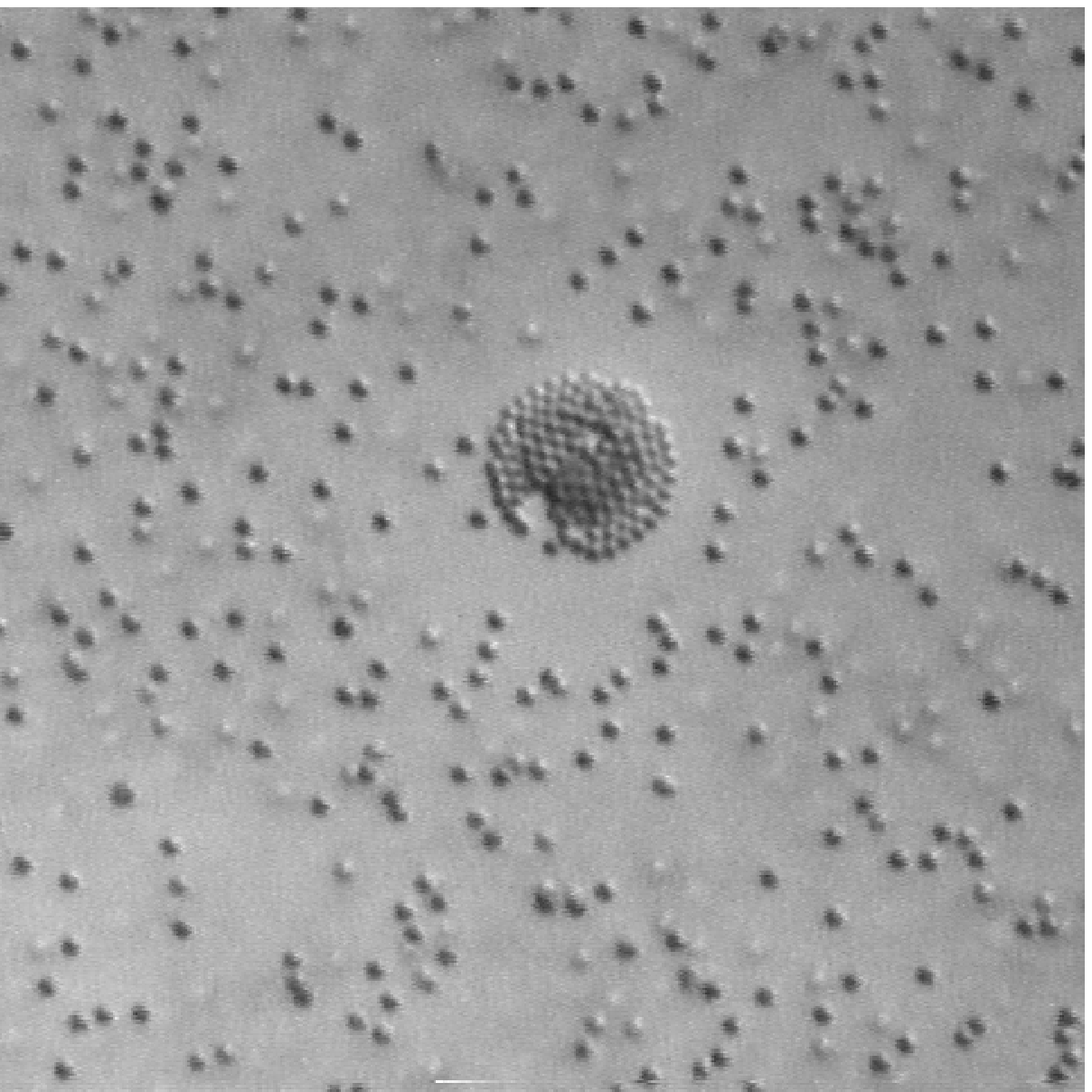}}
\end{minipage}\hfill
\begin{minipage}{3.1truein}
{\epsfysize=3.1truein\epsfbox{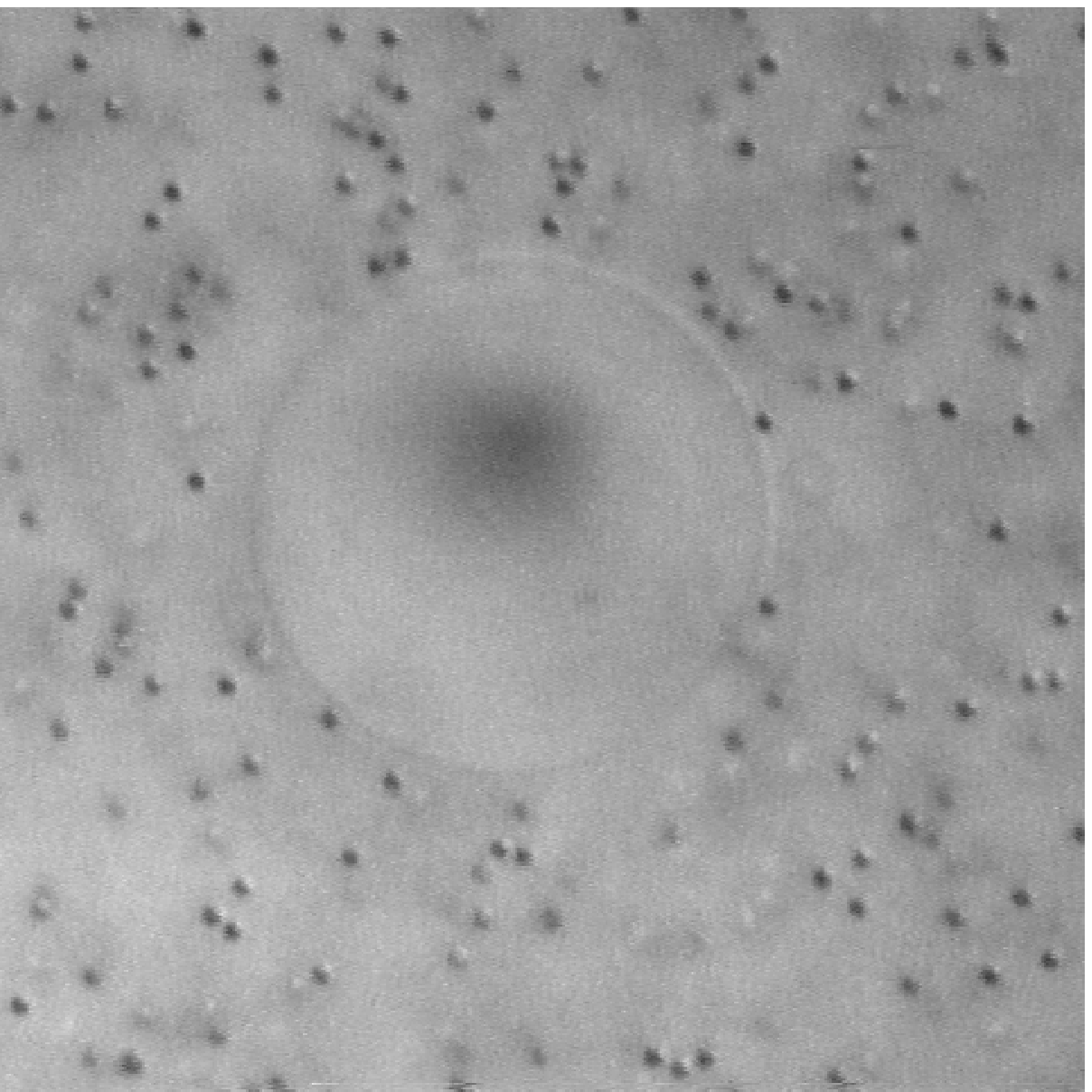}}
\end{minipage}\end{center}
{\caption{{\footnotesize \capphoto}
\label{fphoto}}}\end{figure}
\else
\begin{figure}[h!]\caption{\footnotesize\capphoto \label{fphoto}}\end{figure}
\fi
\fi

The observed behavior is disturbingly at odds with established
intuition and raises several questions:
Why should adhesion to one
zone of the membrane affect adhesion hundreds of screening lengths
away? More urgently, why should electrostatic adhesion saturate in this way?
How can oppositely charged objects repel?

The key to the puzzle is a subtle interplay between the entropic and
electrostatic effects of the mobile counterions and laterally mobile
lipids, which leads to a
thermodynamic instability: The equilibrium state involves the
coexistence of adhesive and repulsive zones in the membrane. The
latter repel incoming negative objects by recruiting
negative counterions on the interior face.
The effects of demixing on membrane
adhesion have been studied
recently by other groups (for example, see
\ifx\ifscience\figtest
\nocite{gelb97a,nard98a}{\it(2,6)}\else\cite{gelb97a,nard98a}\fi). Our
mechanism differs from earlier ones by including cooperative effects
between counterions on {\itppt both sides} of an impermeable membrane.
We will show how this effect can lead to adhesion saturation. A full
discussion will appear elsewhere \cite{dan1}.

\section{Physical picture\label{sphyspic}}
At least three possible equilibrium states could result when a
mixed bilayer vesicle encounters charged surfaces: {\it (i)}~The vesicle
composition could
remain uniform, and thus be uniformly attractive to the approaching
surfaces. In this case, the vesicle should end up
completely covered by particles (or tense and tightly adhering to the
substrate). {\it (ii)}~Alternately, binding could cause total
lateral demixing of the charged and neutral surfactants in the
membrane, and lead to a charge-depleted zone with no attraction to
negative objects. One can easily show, however, that totally
eliminating charged surfactants from the latter zone
comes at a high cost in lateral distribution entropy; instead, enough
residual charge will always
remain to make the depleted zone quite attractive. Thus one would
expect at most {\it (iii)}~a coexistence between high charge
density (tight-adhesion) and low charge density (weak-adhesion) zones.

In the experiments mentioned, however, often none
of the above three expectations was realized: instead, adhesion
saturated at some optimal coverage.
Once this point was reached, the colloidal particles in \cite{ramo1}
are not seen to leave or join the vesicle. Indeed, particles in suspension are
seen to approach, then bounce off, the vesicle. Similarly, the
experiment in \cite{nard98a} found ``blistering'' in the
contact region instead of uniform tight contact.

\def\capone{
{\it (A)~}~Schematic of the
geometry. A vesicle of mixed neutral and cationic surfactants binds
a few anionic objects, then stops, even though
further spheres are available.
\hfil\break
{\it (B)~}~Planar idealization when an approaching charged dielectric
(shaded, above right) is still far away from the
membrane. The membrane interior is at the bottom of the figure. The
zeros denote neutral surfactants, plus signs the charged
surfactants. Circled $\pm$ signs denote counterions in solution. The
solid vertical lines joining
charges are fictitious elastic tethers representing intuitively the
electric field lines; the requirement of charge neutrality translates
visually into the requirement that all charges be tied in this way.
\hfil\break
{\it (C)~}~Redistribution of charges when the negative dielectric
approaches the membrane, if we forbid any charge separation across the
membrane. Four pairs of counterions have been released to infinity
(upper left). The interior monolayer, and its counterion
cloud, are unchanged from {\itppt(B)}. Zone ``b''  presents a net of one
positive charge to the vesicle exterior and so attracts further
incoming negative objects. Dashed horizontal arrows indicate a further
rearrangement of charges allowed once we relax the constraint of zero
charge separation across the membrane.
\hfil\break
{\it (D)~}~The resulting state after the migrations indicated
by dashed arrows in {\itppt(C)}. One additional counterion pair has been
released to infinity and the adhesion gap has narrowed. The net charge
of the bilayer plus
interior counterions in zone ``b'' has reversed sign relative to
{\itppt (C)}.
}

\ifx\ifdontembed\figtest\else
\ifigure{fsetup}{\capone}{%
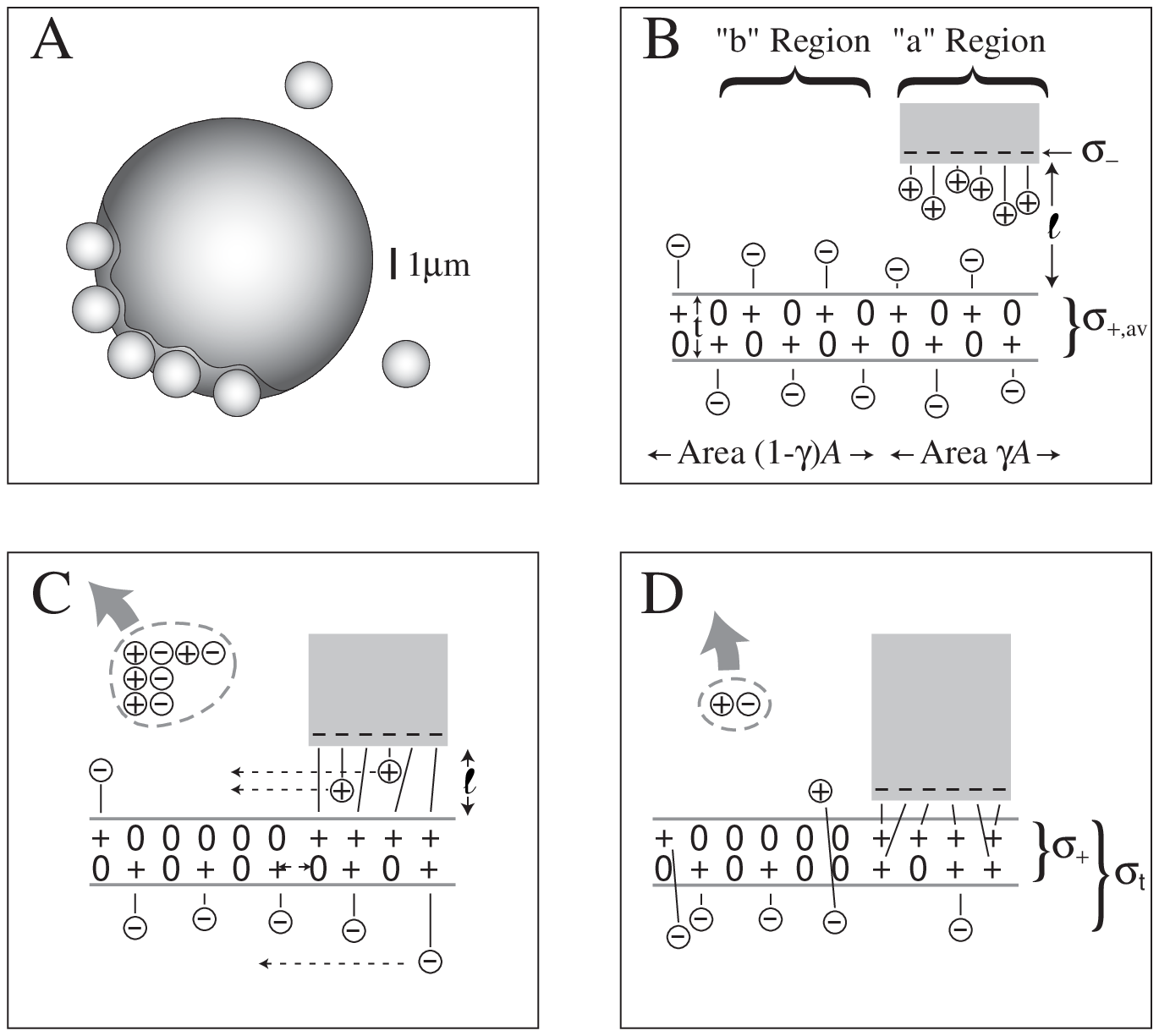
}{4}

\afterpage{\clearpage}\fi

To confront this paradox, we begin with Parsegian and Gingell's
classic analysis of
the attraction of oppositely charged, planar surfaces \cite{pars72a}.
The authors studied the interaction of two infinite parallel planes
with fixed bound surface charge densities $\sigma_+>0$ and
$\sigma_-<0$. Between the planes, a
gap of width $\ell$ contains  water, a dielectric medium with
dielectric constant \cite{fntwo}
$\epsilon=80\epsilon_0$ with mobile point charges (ions) supplied by an
external reservoir. We will  consider all ions to
be univalent, as in the experiment of \cite{ramo1}. The reservoir has
fixed density
$\hat n$ of positive ions and an equal number of negative ions. To either
side of the gap lie infinite dielectrics with no free charges.

In this situation Parsegian and Gingell found that oppositely charged
surfaces initially attract as they are brought in from infinite
separation. The physical mechanism for the attraction is revealing: As
the two surfaces' counterion clouds begin to overlap, a positive
counterion from the negative surface can join a negative counterion
from the positive surface; the pair then escapes to the infinite
reservoir, gaining entropy, without any net separation of charge. The
process continues as the surfaces approach, until one counterion cloud
is completely exhausted. If $|\sigma_-|>\sigma_+$, then at this point
only positive counterions remain in the gap. These residual ions
cannot escape, because that would leave nonzero net charge in the gap
(in the assumed infinite planar geometry, net charge carries an
infinite cost in electric field energy).

At some separation $\ell_*$ the osmotic pressure of the trapped
residual counterions balances the electrostatic attraction of the
plates. Nevertheless, the total free energy change
for bringing the
plates together is always negative: Oppositely charged surfaces always
adhere \cite{pars72a}. This adhesion energy per area is given by
$W\equiv f(\ell_*)-f(\infty) =-(\sigma_+)^2/\epsilon\kappa$, where
$\kappa$ is the
inverse screening length. Remarkably, $W$ is completely
independent of the majority charge density $\sigma_-$
\cite{nard98a}. In light of the
above physical picture, we can readily interpret that fact: The total
counterion release is limited by the {\itppt smaller} of the two
counterion populations.

Thus the physical situation studied by Parsegian and Gingell does not
exhibit adhesion saturation. Fortunately, the general situation we
wish to study differs in three key ways from theirs (see
\tfig{fsetup}, {\itppt A} and {\itppt B}).
\begin{enumerate}\setlength{\itemsep}{0pt}\setlength{\parsep}{0pt}
\item[1)] One of the surfaces is an infinite dielectric of fixed
charge density $\sigma_-$, as above, but the other contains a fluid {\itppt
mixture} of charged and neutral elements. Thus the latter's charge
density $\sigma_+$ may vary, with surface average fixed to some value $\spav$.
\item[2)] The positive surface will be assumed to be a {\itppt membrane}
bounding a closed vesicle of surface area $A$, not the boundary of a
solid dielectric body. The membrane
separates two regions with the same  salt concentration $\hat n$ far
from the membrane.
\item[3)] We will study {\itppt coexistence of two zones} on the membrane: an
attachment zone ``a'' similar to the one studied by Parsegian and Gingell,
and a second zone ``b'', which will eventually prove to be
unattached (we do not {\itppt assume} this).
\end{enumerate}
Because the sizes of the colloidal spheres and the vesicle are much
bigger than the screening length,  our geometry is essentially planar
(\tfig{fsetup}{\itppt B}). (This idealization is self-consistent, as the
equilibrium spacing $\ell_*$ found below will prove to be of order the
screening length.)
For the same reason, we can neglect fringe
fields at the boundaries of the zones ``a'' and ``b''.

Before proceeding with any calculations, we now sketch the new physics
which can arise in the general situation described by points 1 to 3 above.
We will for
concreteness suppose that half the vesicle's counterions, with charge
$A\spav/2$, are
confined to the interior of the vesicle and half to the exterior.

One may be tempted to ignore the interior counterions altogether, in
light of the fact that bilayer membranes are highly impermeable to
ions~\cite{fnthree}. Indeed, counterions trapped inside the vesicle
cannot participate {\itppt directly} in the mechanism described above for
electrostatic adhesion, since they cannot pair with exterior
counterions and escape together to infinity. Accordingly,
let us momentarily suppose that the density of interior
counterions is fixed.

In this situation (fixed interior counterions), Nardi {\itppt
et al.} noted that  zone ``a''
can {\itppt recruit} additional charged surfactants from zone ``b'', in
order to liberate their counterions and improve the adhesion
\cite{nard98a} (\tfig{fsetup}{\itppt B} and {\itppt C}).
The entropic tendency of the charged and uncharged surfactants to
remain mixed opposes this redistribution, however, and the resulting
adhesion is a compromise between the two effects.
Zone ``b'' will still have nonnegative
charge, and will still remain quite attractive to further colloidal
particles; there is no adhesion saturation.

The argument in the previous paragraph, however, neglects the ability
of interior counterions to move {\itppt laterally}. As
shown in \tfig{fsetup}{\itppt C}, the approaching exterior negative
object will push negative interior counterions out of zone ``a'' and
into zone ``b'', where they can overwhelm the residual positive
membrane charge and effectively reverse its sign.

This rearrangement liberates {\itppt exterior}
counterions from both zones as shown in \tfig{fsetup}{\itppt D},
enhancing the adhesion. The capacitive energy cost of separating
charge across a membrane in this way is
significant, because of the low dielectric constant
$\epsilon_m\approx2\epsilon_0$ of the hydrocarbon tails of lipids
and other surfactants. Nevertheless, the cost is
{\itppt initially} zero, being proportional to the {\itppt square} of the
charge separated, and hence there will always be some lateral
rearrangement inside the vesicle, as indicated by the dashed arrows in
\tfig{fsetup}{\itppt C};
\tfig{fsetup}{\itppt D} is the result. If this rearrangement reverses the
effective membrane charge, it will lead to active repulsion in the
nonadhesion ``b'' zone.

We will now show that the scenario just sketched can actually occur
under a broad range of experimentally-realizable conditions.

\section{Calculations\label{scalculations}}
Consider the coexistence of two homogeneous zones
``a'' and ``b'' with area $\gamma A$ and $(1-\gamma)A$ respectively.
We will
for simplicity assume that all surfactants have the same fixed area
per headgroup $a_0$ and charge either $+e$ or 0. We must
compute the equilibrium value $\gamma_*$ of the fractional area coverage in
terms of
the ambient salt concentration $\hat n$, the dielectric charge density
$\sigma_-$,
the average membrane composition $\spav$, and the
headgroup area $a_0$. We will show that the effective charge in
zone ``b'' is negative.

Examining \tfig{fsetup}, we see that each zone freely exchanges {\itppt
two} independent conserved quantities with the other. We may take these to be
the
net counterion charge $Q_1$ below the membrane and the total
surfactant charge $Q_+$ of the membrane itself, with corresponding
areal charge densities $\sigma_1$ and $\sigma_+$ respectively
\cite{fnfour}. We
express all densities in dimensionless form, letting $\smax=2e/a_0$ and
$\sob=\sigma_1/\smax$, {\itppt etc.} Thus $\spb$ must obey the important
conditions $0<\spb<1$, while $\sob$ is in principle unbounded.

\subsection{Thin membrane limit\label{ssthincalc}}
To make the formul\ae{} as transparent as possible, we first study the
hypothetical case of a very thin membrane. We must
compute the free energy density of
a homogeneous region at fixed charge density, and then apply the usual phase
coexistence rules. To get $f$, we simply add three terms, letting
$f=f_1+f_0+\fm$, where
\begin{enumerate}\item[1)]
$f_1$ is the free energy of the half-infinite space inside the
vesicle. This space
sees a plane of charge density $-\sigma_1$, so in Debye-H\"uckel
theory its free
energy cost is $f_1=({\smax}^2/2\kappa\epsilon){\sob}^{\,2}$.
\item[2)] $f_0$ is the free energy of the gap region. This space
sees a plane of charge $\sigma_-$, a gap of width $\ell$, and another
plane of total charge $\sigma_{\rm
t}\equiv\sigma_++\sigma_1$. Minimizing the free
energy over $\ell$ gives in Debye-H\"uckel theory \cite{pars72a}
$
f_0={{\smax}^2\over2\kappa\epsilon}\left[
{\smb}^{\,2}+{\sob}^{\,2}+{\stb}^{\,2}+\bar W
\right]
$
. As discussed above, the nondimensional adhesion energy $\bar W$ equals
$-2{\smb}^{\,2}	$ if $\stb>|\smb|$,
$-2{\stb}^{\,2}	$ if $0<\stb<|\smb|$, or
zero if $\stb<0$.
The third case
corresponds to the possibility of a  charge-reversed state
with equilibrium spacing $\ell_*=\infty$ (zone ``b'' of \tfig{fsetup}{\itppt
D}).
\item[3)] $\fm$ is the free energy density of the membrane itself. We
retain only the entropy of mixing of charged and neutral surfactants, and
neglect any other entropic or enthalpic packing effects in the
membrane's free energy. Thus, we have the simple form
$
\fm={2\over a_0}\kbt\left[
\spb\log\spb + (1-\sbp)\log(1-\sbp)\right]
$
.\end{enumerate}
The mixing entropy term $\fm$  opposes phase decomposition, whereas
the electrostatic terms
$f_1+f_0$ promote it. The dimensionless ratio
$\beta\equiv2\kappa\epsilon\kbt/e\smax$ describes the relative
importance of these effects.
Because typical surfactants have $\smax=e/0.6\,$nm$^2$, a
1~mM NaCl solution with $\kappa^{-1}\approx10\,$nm gives
$\beta\approx0.006$. We may thus expect to find
two-phase coexistence and indeed inspection of the free energy density
reveals such an instability (\tfig{f3dthin}). (In the figures we have
plotted the exact Poisson-Boltzmann theory result \cite{dan1};
these results are qualitatively similar
to those derived from the simple, linearized Debye-H\"uckel
formul\ae{} given above \cite{fnone}.)

\def\capfthreedthin{Free energy density  for a
fictitious,
infinitely thin membrane. The variables $\stb$ and $\spb$ are
nondimensional forms of the charge densities $\sigma_+$ and $\sigma_{\rm t}$ in
\tfig{fsetup}{\itppt D}.
For
easier visualization we have inverted the figure, rescaled, and added
a linear function, plotting \offsetthreed{} instead of $f$.
For illustration we
have taken the average membrane composition $\spavb=0.5$
(corresponding to 1:1 mole fraction of charged
surfactant), the normalized colloidal particle charge
$|\smb|=1.5$, and $\beta=0.006$ (see text). The solid curve
is the locus of points where $\smb=0$; points to the left of this curve
represent charge-reversed states.
The geometric fact that the surface shown has two hills on opposite
sides of the solid curve implies that
the system's ground state consists of two coexisting zones, one of which is
charge-reversed.
The common-tangent construction from thermodynamics determines the
composition of the two zones. }

\ifx\ifdontembed\figtest\else
\ifigure{f3dthin}{\capfthreedthin
}{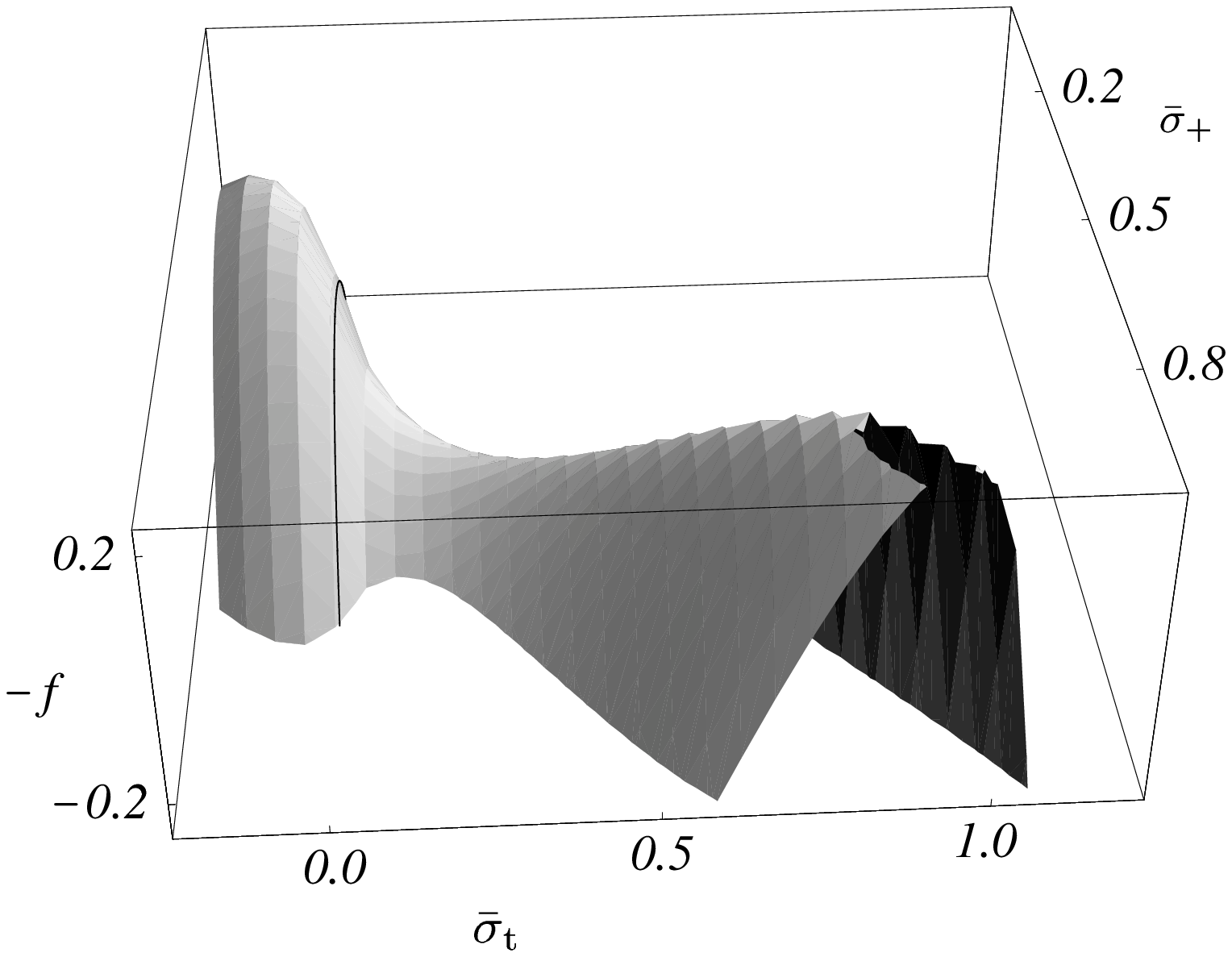
}{3}
\fi

Using for illustration the values $\spavb=0.5$ and $\smb=-1.5$ then gives
\cite{dan1} coexistence between
an adhesion zone with $\spb^{(a)}=0.95$, covering a fraction $\gamma_*=36$\%
of the vesicle, and a charge-reversed zone with $\spb^{(b)}=0.25$. The latter
zone presents total charge density $\stb=-0.12$ to the outside of
the vesicle (thus reversed in sign),  or about $-45$\% of the value $\spav/2$
presented to the outside when the adhering dielectric is far away.

\subsection{Realistic membrane\label{ssrealcalc}}
To treat a realistic (finite-thickness) membrane, we must distinguish the two
halves of the bilayer. Intuitively, one may expect that for a  very
thick membrane the energy cost of putting electric field lines in the
dielectric interior of the membrane would become prohibitive, so that
the system stops at \tfig{fsetup}{\itppt C}\/ instead of proceeding to
\tfig{fsetup}{\itppt D}. We now show that realistic membranes are {\itppt not}
so thick,
and do exhibit the same charge reversal (to a reduced degree) as the thin
case just discussed.

Let the inner monolayer have charge fraction
$u\spb$ and the outer $(1-u)\spb$.  Bilayer membranes have
capacitance per area of around $c=0.01\,$pF/$\mu$m$^2$ \cite{genn89a},
so  we modify our free energy density $f$ by adding a
capacitive term $f_{\rm c}={{\smax}^2\over2\kappa\epsilon}
\tau\bigl(\sob+u\spb\bigr)^2$, where the dimensionless ratio
$\tau\equiv\kappa\epsilon/c$ measures the importance of membrane thickness.
Using for illustration a 1~mM NaCl
electrolyte then gives $\tau\approx\tauchoice$. We must also replace
$\fm$ by the corresponding formula for {\itppt two} layers.

For given $(\spb,\stb)$, we first minimize $f(\spb,\stb,u,\ell)$ over
$\ell$ and $u$, then repeat the phase-coexistence analysis.
The free energy surface is then qualitatively similar to \tfig{f3dthin},
though the extent of coverage in equilibrium $\gamma_*$ is larger,
around 63\% \cite{fnX}. The degree of charge-reversal
is now smaller, about $-1.2$\% of the value $\spav/2$. Even this
small effect causes
vigorous rejection of additional adhering objects: Increasing the
adhesion area beyond its preferred value $A\gamma_*$ by 1$\,\mu$m$^2$
on a vesicle of radius 10$\,\mu$m
comes at a net free energy cost of more than $3000\kbt$. Decreasing the
area below $A\gamma_*$ by removing a ball comes at a similar cost.
Thus, realistic membranes can partition into an adhesion
zone and a charge-reversed, repulsive, zone.

Our result is relatively insensitive to the values of the charge
densities $\sigma_-$ and $\spav$, though we must have
$|\sigma_-|>\spav/2$ in order to obtain the instability. Increasing the
salt concentration beyond $\nhat=20\,$mM, however, eliminates charge reversal
by increasing $\tau$, a prediction in qualitative agreement with the
experiments of \cite{ramo1}. If $\nhat$ lies between 20~mM and about
150~mM, we still find an instability, this time to partitioning into
strong- and weak-adhesion zones \cite{nard98a}.

\ifx\ifscience\else\section{Conclusion}
We have shown how a positively charged, mixed vesicle can develop
charge-reversed
zones which {\itppt repel} oppositely charged surfaces and particles,
via the mechanism shown in \tfig{fsetup}. We began with Nardi {\itppt et
al.}'s observation that mixed membranes can be unstable to spontaneous
lateral segregation of charged and neutral surfactants, then added the
possibility of a concomitant segregation of interior counterions. \fi{}In
retrospect our mechanism is reminiscent of the chemiosmotic principle
in bioenergetics \cite{nossalbook}: In this context it is well known
that electrostatic
effects can be transmitted over many screening lengths with the help
of a semipermeable membrane.
Besides entering into an explanation of the experiments in
\cite{nard98a,ramo1}, our mechanism predicts that flaccid charged vesicles can
adhere to oppositely charged substrates while remaining flaccid. Our analysis
also makes testable predictions about the dependence of the
equilibrium area fraction $\gamma_*$ on the system parameters, notably
the bilayer composition and salt concentration. Perhaps most
strikingly, the charge-reversed zone
found here should prove {\itppt attractive} to {\itppt same-charge} objects
--- a phenomenon not yet seen.

\def\ackstext{
We  thank R. Bruinsma and S. Safran for
discussions, J. Crocker, K. Krishana, and E. Weeks for experimental
assistance, and J. Nardi for communicating his results
to us before publication.
ND was supported in part by NSF grant CTS-9814398;
TCL, LR, and DAW were supported in part by  NSF Materials Research and
Engineering Center Program under award number DMR96-32598 and equipment
grants DMR97--04300 and DMR97-24486;
PN was supported in part by NSF grant DMR98-07156;
LR was supported in part by a Bourse Lavoisier from the Minist\`ere
des affaires Etrang\`eres de France.}
\ifx\ifscience\figtest\else
\paragraph{\bf Acknowledgments}
\ackstext
\fi

\newpage
\ifx\ifscience\figtest\bibliographystyle{prsty}\else
\bibliographystyle{unsrt}\fi
\bibliography{repel}

\begin{thebibliography}{10}
\bibitem{fnY}
Changes in the materials used had little effect on the experimental results in
  \cite{ramo1}. We used colloidal spheres of diameters 0.98, 0.83 and
  1.00$\,\mu$m, with sulfate groups, sulfate and carboxylate groups, and
  carboxylate and amine groups respectively. Electrophoretic velocity
  measurements confirmed that all types of particles were negatively charged in
  our experimental conditions. All three samples behaved similarly in the
  experiments. We also varied the bilayer characteristics through the addition
  of octanol. Although this co-surfactant is known to alter the
  curvature-elasticity properties of bilayers, we could not see any qualitative
  effect on our results at alcohol to Triton-X weight ratios ranging from 0 to
  0.7. In \tfig{fphoto} the weight ratio DDAB:Triton:octanol was 1:0.37:0.13.
  The buffer solution contained NaCl at concentrations from 1 to 50~mM, as
  discussed below.
\bibitem{fntwo}
{We work in SI units. Thus $\epsilon_0=9\eem{12}\,$Farad/meter, the potential
  around a point charge $q$ in vacuum is $\psi(r)=q/4\pi\epsilon_0 r$, {\itppt
  etc.}}
\bibitem{fnthree}
We need not assume the membrane to be impermeable to {\itppt water}; because
  the bulk salt concentration is assumed the same on both sides, there will be
  no net osmotic flow.
\bibitem{fnfour}
A third exchanged quantity, the net counterion charge density {\itppt outside}
  the membrane, is then fixed by charge neutrality:
  $\sigma_0=-(\sigma_1+\sigma_+ +\sigma_-)$. Similarly the density of neutral
  surfactants in the membrane is not independent, being given by
  $(2/a_0)-(\sigma_+/e)$. The numbers of individual counterions of each species
  are {\itppt not}, however, conserved, since neutral $\pm$ pairs can be
  exchanged with large reservoirs (the bulk solution inside and outside the
  vesicle) without macroscopic charge separation.
\bibitem{fnone}
Poisson-Boltzmann theory is a mean-field approximation. It is well known that
  correlated fluctuations can cause surprising effects if the counterions are
  multivalent. For example, like-charged parallel plates can have attractive
  regimes \cite{guld84a,rouz96a}, though we are not aware of a similar
  prediction of repulsion for opposite charges. Correlation effects are
  generally small if the ions are univalent, as in the experiment of
  \cite{ramo1}. The physical mechanism presented here, in contrast, arises
  already in mean field theory; it does not rely on any fluctuation effects
  (nor on finite ion-size effects). Including such effects does not materially
  change our conclusions \cite{dan1}. Other colloidal forces not included in
  our analysis (for example solvation forces) are short-ranged; our effect
  relies on the physics of surfaces when they are separated by more than a
  nanometer.
\bibitem{fnX}
The extent of coverage is not directly measurable, since the degree to which
  each colloidal sphere is engulfed by membrane is not observed. In addition,
  the actual situation in \tfig{fphoto} may not be in equilibrium. Our point is
  simply that our theory explains why the vesicle in \tfig{fphoto} has stopped
  attracting further colloidal particles.
\end{thebibliography}

\ifx\ifscience\figtest\bigskip
\ackstext
\fi

\newpage

\ifx\ifdontembed\figtest
\begin{center}{\bf Figure Captions}\end{center}

\bigskip
\tfig{fphoto}. {\capphoto}

\bigskip
\tfig{fsetup}. {\capone}

\bigskip
\tfig{f3dthin}. {\capfthreedthin}

\ifx\figsin\figtest
\newpage
\begin{figure}[h!]
\begin{center}
\begin{minipage}{3.1truein}
{\epsfysize=3.1truein\epsfbox{/u/nelson/Ms/Rafts/PixPhotoShop/raft1a.eps}}
\end{minipage}\hfill
\begin{minipage}{3.1truein}
{\epsfysize=3.1truein\epsfbox{/u/nelson/Ms/Rafts/PixPhotoShop/raft1b.eps}}
\end{minipage}\end{center}
{\caption{
\label{fphoto}}}\end{figure}

\newpage
\ifigure{fsetup}{}{%
oldfig2.eps%
}{4}

\ifigure{f3dthin}{}{fig3xfig.eps
}{3}

\else
\newpage\begin{figure}\caption{
\label{fphoto}}\end{figure}
\begin{figure}\caption{\label{fsetup}}\end{figure}
\begin{figure}\caption{\label{f3dthin}}\end{figure}
\fi  
\fi  

\end{document}